\documentclass[review]{elsarticle}
\usepackage{amsmath}
\usepackage{float}
\usepackage{xcolor}
\usepackage{bbm}
\usepackage{amssymb}
\usepackage{tikz}
\usepackage{verbatim}
\usetikzlibrary{calc}
\usetikzlibrary{shapes,arrows}
\usepackage{caption}
\usepackage{makecell}
\usepackage{subcaption}
\usepackage{mathptmx}

\begin{document}
\begin{frontmatter}
\title{Surrogate Modeling of Stochastic Functions - Application to computational Electromagnetic Dosimetry}


\author[mymainaddress]{Soumaya Azzi\corref{mycorrespondingauthor}}
\cortext[mycorrespondingauthor]{Corresponding author}
\ead{soumaya.azzi@telecom-paristech.fr}

\author[mymainaddress]{Yuanyuan Huang}

\author[mysecondaryaddress]{Bruno Sudret}

\author[mymainaddress]{Joe Wiart}

\address[mymainaddress]{Telecom ParisTech, Chair Mod\'{e}lisation, Caract\'{e}risation et Ma\^{\i}trise, 46 Rue Barrault, 75013 Paris}
\address[mysecondaryaddress]{ETH Z\"{u}rich, Chair of Risk, Safety and Uncertainty Quantification, Stefano-Franscini-Platz 5, 8093 Z\"{u}rich, Switzerland}

\begin{abstract}
Metamodeling of complex numerical systems has recently attracted the interest of the mathematical programming community. Despite the progress in high performance computing, simulations remain costly,  as a matter of fact, the assessment of the  exposure to radio frequency electromagnetic fields is  computationally prohibitive since one simulation can require hours.
Moreover, in many engineering problems, carrying out deterministic numerical operations without considering uncertainties can lead to unreliable designs. In this paper we focus on the surrogate modeling of a particular type of computational models called stochastic simulators. In contrast to deterministic simulators which yield a unique output for each set of input parameters, stochastic simulators inherently contain some sources of randomness and the output at a given point is a probability density function. Characterizing the stochastic simulators is even more time consuming. This paper represents stochastic simulators as a stochastic process and describes a metamodeling approach based on the Karhunen-Lo\`eve spectral decomposition.
\end{abstract}


\end{frontmatter}


\section{Introduction}

In parallel with the widespread use of wireless systems, an increased risk perception related to radio-frequency electromagnetic fields (RF-EMF) has been observed \cite{Eurobarometer}, and the assessment of the human exposure to RF-EMF has aroused social attention. To respond to such concerns, large efforts have been carried out to establish methods to verify compliance with exposure limits.
The human EMF exposure can be quantified in terms of Specific Absorption Rate (SAR) expressed in W/kg and representing the RF power absorbed per unit of mass of biological tissues. 

As a matter of fact, the RF-EMF sources are the combination of uplink and downlink radiations coming from, respectively, personal wireless devices (e.g. smartphones or tablets) and cellular base stations (BS) or access points \cite{Huang2017}.  
In this respect, advanced computational propagation tools were used in many studies \cite{Huang2017,Varsier2015,Huang2016} to characterize the signal's attenuation between  a transmitter and  a receiver. Such tools can provide accurate path loss results, however they are strongly dependent on detailed building and terrain data, which increases the computational burden.
To overcome such a limitation, it is proposed in \cite{Huang2017} to decrease the calculation effort using the \textit{path loss exponent} (PLE). The PLE is a positive number characterizing
the attenuation of the power received by the device relative to what has been emitted by the base station.
In this paper the PLEs are derived from the path loss calculation based on random cities having the same macroscopic structures, such as street width, building height or street angle. The assessment of PLE over a city can be seen as a random function of this city's morphological features. The characterization of the probability density function (PDF) of one PLE requires a large number of calculations. 
Taking into account the important computational cost for one PLE calculation (i.e., more than one hour by means of a computer type Intel Xeon E5-2620V3 2.4GHz 6Core 15Mo and NVIDIA TESLA K80), a surrogate model is needed.

Over the last decade, large efforts have been made to build surrogate models of deterministic functions. The most popular are Gaussian process modeling (a.k.a Kriging) \cite{Santner2003}, generalized polynomial chaos expansion GPCE \cite{Ghanembook2003,Xiu2002} and low rank tensor approximations \cite{Chevreuil2015, KonakliJCP2016, Chiaramello}. Metamodeling of stochastic functions is a less mature field.  Assuming that the model output is a Gaussian field trajectory, recent studies \cite{Browne2016,Marrel2012,Ankenman2009,Bursztyn2006} build two independent or joint deterministic metamodels to fit the mean and the covariance of the assumed Gaussian process. 
Also based on the joint metamodeling approach, \cite{Iooss2009} simultaneously surrogates the mean and the dispersion using two interlinked Generalized Additive Models. Alternatively, the study carried out in  \cite{Moutoussamy2015} focused on projecting the output density on a basis of chosen probability density functions. With this approach, the coefficients are computed  by solving constraint optimization problems for the purpose of building a local metamodel. This method is not ideal for assessing certain quantities of interest (e.g., quantiles).

To overcome these limitations, this paper describes a non-parametric method, based on the Karhunen-Lo\`eve (KL) decomposition, to build a surrogate model of random functions.
The paper is organized as follows: In Section 2, the EMF exposure assessment methodology is introduced and the assessment of the PLE in stochastic cities is presented. Section 3 briefly summarizes the Kriging surrogate model and the polynomial chaos expansion, then introduces the proposed approach which makes use of the Kl spectral decomposition. Two applications are illustrated in Section 4. Discussion and conclusions are provided in section 5.

\section{Simplified assessment of RF-EMF exposure}

\subsection{Simplified assessment of RF exposure using a path loss exponent}
 

To date, various advanced computational tools are used to design telecommunication infrastructures. Briefly, these tools use radio channel models to predict the narrow-band path loss and thus the radio coverage across a specific environment. 
To do so, some ray-based physical propagation mechanisms such as reflections and diffractions are implemented in the channel models. This so-called 'ray tracing' technique depends on the digital geographical map data extracted from the real environment, allowing for an accurate estimation of the path loss between the base station antenna and the wireless device \cite{YLostanlen}. The emitted and received power can then be estimated and used to evaluate the EMF exposure. A limit of such a technique is the very high computational cost due to the use of complex deterministic propagation models. A simplified path loss model was therefore proposed in \cite{Huang2017} for the assessment of both uplink and downlink radio emissions. With such an approach, the received power is modeled as a function of the propagation distance weighted by a PLE.
Thus, the received power $P$ can be modeled as follows:
\begin{equation}
P(d)=c-10 \,\alpha\, \log_{10}(d)
\label{equ:equation1}
\end{equation}
where $d$ is the distance between the transmitter and the receiver, $\alpha$ is the PLE, and $c$ is a constant parameter. All these parameters can be estimated using ordinary least-squares method.

As explained previously, the PLE highly depends on the features governing the city structure, such as the organization of buildings into blocks, the street intersections and  the street network anisotropy.
To cover the variability of PLE which might be observed in the same kind of cities, random city samples have to be generated.

\subsection{PLE statistical distribution assessment using stochastic cities }

\begin{figure}[!t]
	\centering
	\begin{tabular}{cc}
		
		\includegraphics[scale=0.5]{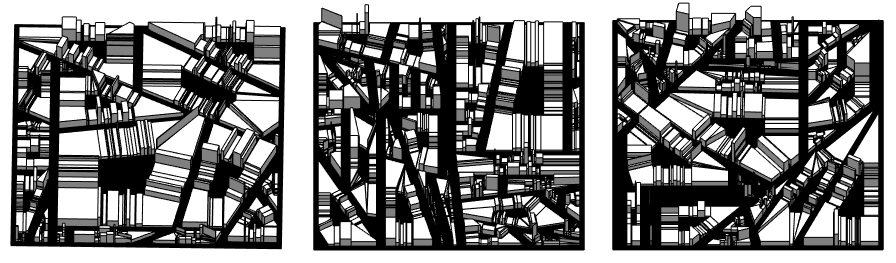} 
		
	\end{tabular}
	\caption{Examples of 3D stochastic city models with same values of morphological features.}
	\label{fig:fig1}
\end{figure}

Stochastic geometry has proven its ability to describe the complex structures of a city \cite{Courtat2016} via a limited number of parameters, such as building density, street width, number of intersections, etc. Based on statistical distributions of the city features, i.e., building height, street width, anisotropy, a stochastic geometry simulator \cite{Courtat2016} was used to generate various random 3D cities. The inputs of this stochastic generator are the mean values of those distributions. Figure \ref{fig:fig1} illustrates various city samples generated with the same morphological features.

Thereafter, the EM attenuation map related to each city sample was obtained through a 3-D ray launching technique \cite{YLostanlen}, which is commonly used to propagate EMF in urban areas.
In a given realization of a stochastic city, an antenna operating at 2600 MHz and having a total emitted power $P_0$ has been located in the city and millions of rays have been launched. The signal attenuation map can thus be obtained by assessing the received power in the 'measurement' plane (1.5 m above the ground). Different stochastic city realizations, based on the same set of city parameters, can have different signal attenuation maps. For each of these realizations, using ordinary least-square approximation and Eq. (\ref{equ:equation1}), the mean value of the PLE is obtained. This value is considered  as a realization of a random variable having a probability density function that depends on the input features of the city. As a consequence, the model can be considered as a stochastic process indexed by the city parameters. The assessment of the PLE distribution requires huge computational time. It is therefore of interest to build a surrogate model for such a stochastic simulator.



\section{Surrogate modeling of random functions}
\subsection{Meta-modeling of deterministic functions}

Despite the progress in high performance calculation, simulations still require a large computation time. 
Therefore, uncertainty quantification cannot be carried out using classical approaches such as Monte Carlo simulation. To overcome this limitation, surrogate models (a.k.a. metamodels) have been studied to emulate the output of a complex computational model. Of interest for us are the so-called polynomial chaos expansion and Kriging. The two techniques are now briefly reviewed for the sake of completeness. 
\subsubsection*{Polynomial chaos expansion}
 Polynomial chaos expansions (PCE) consist in expanding the random output onto an orthogonal basis $\{\Psi_{\boldsymbol{\beta}}, \boldsymbol{\beta} \in \mathbb{N}^d \}$ with respect to the joint probability density function (PDF) of the input parameters \cite{Ghanembook2003} \cite{SudretHDR}.

Consider a probability space $(\Omega, \mathcal{F}, \mathbb{P})$, where $\Omega$ is the event space, $\mathcal{F}$ its $\sigma$-algebra and $\mathbb{P}$ its probability measure. 
Consider a numerical model $H$ with independent input parameters gathered in a random vector $\textbf{X}$ of size $d$ with a joint probability density function $f_\textbf{X}$. Suppose the random output has a finite variance, i.e. $\mathbb{E} \left[  H(\textbf{X}) \right]  ^2  <\infty$. $H(\textbf{X})$  can be expressed as follows :
\begin{equation}
 H(\textbf{X})=  \sum\limits_{ \boldsymbol{\beta} \in \mathbb{N}^d}  {a}_{\boldsymbol{\beta}} \Psi_{\boldsymbol{\beta}} (\textbf{X})
 \end{equation}
where $a_{\boldsymbol{\beta}}$ are unknown deterministic coefficients and $\Psi_{\boldsymbol{\beta}}$ are multivariate polynomials obtained as tensor products of univariate polynomials of degree $(\beta_1,\dots, \beta_d)$

To determine the coefficients $a_{\boldsymbol{\beta}}$  there are two main-stream methods, either using   projection methods \cite{XiuBook2010} where the expansion is projected onto the polynomial space, or by casting a least-squares minimization problem \cite{Berveiller2006,BlatmanCras2008}. A nice feature of PCE is the simplicity with which one obtains the most used statistics of the quantities of interest: mean, variance as well as Sobol sensitivity indices \cite{Sobol1993,Saltelli2004} 
can be computed analytically from the estimated coefficients \cite{SudretRESS2008b}.

\subsubsection*{Kriging} 
Kriging (a.k.a Gaussian process modelling)  starts with a prior distribution over the output $H(\boldsymbol{x})$. It treats the deterministic response of $H(\boldsymbol{x})$ as a realization $\mathcal{F}(\boldsymbol{x},\omega), \omega \in \Omega$ of a stochastic process $\mathcal{F}(\boldsymbol{x})$ such as:
\begin{equation}
 \mathcal{F}(\boldsymbol{x})= \mu(\boldsymbol{x}) + Z(\boldsymbol{x})
\end{equation}
where $\mu(\boldsymbol{x})$ is the global model mean. $Z(\boldsymbol{x})$ is assumed to be a zero-mean Gaussian random process with the following properties: 

\begin{equation}
\mathbb{E}\left[ Z(\boldsymbol{x})\right] =0 ~~~~~~
\text{Cov}[Z(\boldsymbol{x}),Z(\boldsymbol{x}')]=\sigma^{2}k(\boldsymbol{x},\boldsymbol{x}')
\end{equation}
where $\sigma^{2}$ is the process variance and $k(\boldsymbol{x},\boldsymbol{x}')$ is the correlation function between any two locations $\boldsymbol{x}$ and $\boldsymbol{x}'$. $k(\boldsymbol{x},\boldsymbol{x}')$ is defined as a function of the Euclidean distance $h = \lVert \boldsymbol{x}-\boldsymbol{x}'\rVert_2 $ with a set of constants called hyperparameters $\theta$. $k(\boldsymbol{x},\boldsymbol{x}')$, which is a correlation function between any two points $\boldsymbol{x}$ and $\boldsymbol{x}'$ of the input parameter space, is represented for instance as a product of univariate correlation functions for each variable as follows:
\begin{equation}
k(\boldsymbol{x},\boldsymbol{x}')= \prod_{i=1}^n k(x_i,x_i') ~~~
\text{or as:} ~~~ k(\boldsymbol{x},\boldsymbol{x}')= R(h), ~~ h= \sqrt{\sum\limits_{i=1}^{n} \left(  \frac{x_i-x_i^{'}}{\theta_{i}}\right) ^2}
\end{equation}
Standard correlation functions (also known as \textit{kernels}) are the Gaussian, exponential and Matern kernels \cite{Santner2003}.

Depending on the stochastic properties of the random field and the various degrees of stationarity assumed, different methods for calculating the hyperparameters of $k$ can be deduced \cite{Santner2003}.

 \subsection{Surrogate modeling of random functions }
 Consider a probability space $(\Omega, \mathcal{F}, \mathbb{P})$ where $\Omega$ is a sample space, $\mathcal{F}$ is a $\sigma$-algebra and $\mathbb{P}$ the probability measure. $H_{\omega}(\boldsymbol{x})=H(\boldsymbol{x}, \omega),~ \omega \in \Omega$ denotes a stochastic model, in other words,  a stochastic process indexed by $\boldsymbol{x} \in D$, where $D \subset \mathbb{R}^n$ is the  domain of definition of $\boldsymbol{x}$.

 A stochastic process (stochastic function), such as the PLE described previously, is a family of random variables $\{ {H}_{\boldsymbol{x}}, \boldsymbol{x} \in \mathcal{D} \}$ defined on the same probability space. At a fixed $\boldsymbol{x}$, $ H(\boldsymbol{x},\omega)$ is a random variable, for a fixed $\omega$, $H(\boldsymbol{x},\omega)$ is a deterministic function of $\boldsymbol{x}$ and is called a trajectory. The covariance function of the process reads as follow:
 \begin{equation}
     C(\boldsymbol{x},\boldsymbol{y})=\mathbb{E}[H(\boldsymbol{x},\omega) H(\boldsymbol{y},\omega) ]
\label{equ:equation2}
\end{equation}

The simulation tools used in this paper make it possible to 'freeze' the randomness $\omega$ and hence simulate trajectories by sampling, \textit{for a frozen $\omega$}, the model response at different values of $\boldsymbol{x}$. In other words, we are able to generate $H(\boldsymbol{x}^{(1)},w_k)$ and  $H(\boldsymbol{x}^{(2)},w_k)$ with the same $w_k$, where $w_k$ is an internal source of stochasticity in the simulation tool. This assumption enables us to compute the empirical covariance function of the model output, and thus apply the KL decomposition that can be used to model the random process \cite{Ghanembook2003}. In the following, we first present the decomposition in more details, then we explain how the decomposition is used to surrogate a stochastic process.

\subsection{Metamodeling of random function based on KL expansion} 

Surrogate models allow one to build an approximate model $\widehat{H}$ such that $ H \simeq \widehat{H}$. As pointed out in the introduction the computational cost of the simulations is significant, the  objective is therefore to have a parsimonious approach. 
To metamodel stochastic functions we will consider a random function that can be assessed over points belonging to a Design of Experiment $\text{DoE}=\left\lbrace \boldsymbol{x}^{\left( 1\right) },\dots , \boldsymbol{x}^{\left( M \right)}  \right\rbrace  \subset D$. As pointed out we can simulate the process in different points with the same random seed. For each point of $DoE$ ($M$ points), simulations have been carried out using $N$ different random seeds, which corresponds to generate $N$ trajectories of the random process at $M$ discrete points. Using KL decomposition, the random process can be modeled as \cite{Ghanembook2003}: 
\begin{equation}
    H(\boldsymbol{x},\omega) \simeq \sum_{i=1}^{M} \sqrt{\lambda_i} \xi_i(\omega) \phi_i (\boldsymbol{x})
\label{equ:equation6}
\end{equation}
where $\boldsymbol{\phi_i}$ and $\lambda_i$ are respectively the eigenvectors and the eigenvalues of $C(\boldsymbol{x},\boldsymbol{y})$ . $\xi_i$  are uncorrelated random variables with unit variance (detailed in Section \ref{Random variables}) and are given by, 
\begin{equation}
 \xi_i(\omega)= \dfrac{1}{\sqrt{\lambda_i}} \int_D H(x,\omega) \phi_i(x) dx
 \label{equ:equation5}
 \end{equation}
 
Eq. (\ref{equ:equation6}) requires the knowledge of $ \xi_{i} (\omega)$ and $\phi_{i}(\boldsymbol{x}) $  for all $\boldsymbol{x} \in D $. However the eigenvectors  $\phi_{i}(\boldsymbol{x}) $ are only known at the discrete sample points of the DoE.
In order to get the value of the eigenvectors over the domain of interest, we proceed in two different ways:
\begin{itemize}
 \item  metamodeling the eigenvectors via usual surrogate modeling methods;\\
 \item  or surrogating the empirical covariance, then find the new eigenvectors on the domain of interest.  \\
 \end{itemize}
As far as the random variables $ \xi_{i} (\omega)$ are concerned, they can be obtained as the projection of the random process over the $\phi_{i}(\boldsymbol{x}) $ (see Section \ref{Random variables}) .
 
 The following sections describe the two approaches as well as the way the random variables $\xi_{i}(\omega)$ are characterized.  
 
\subsubsection{Eigenvectors interpolation}
 A first step to get the eigenvectors all over $D$ would be to interpolate the eigenvectors $ \boldsymbol{\phi}_i $. The KL expansion (Eq. (\ref{equ:equation6})) will then read as follow: 
\begin{equation}
    H(\boldsymbol{x^*},w)=\sum_{i=1}^{M} \sqrt{\lambda_i} \xi_i(\omega) \hat{\phi_i} (\boldsymbol{x^*})
\label{equ:equation7}
\end{equation}
where $ \boldsymbol{\hat{\phi}}_i$ is an approximation of the true eigenvector $\boldsymbol{\phi}_i (.), ~ i = 1, \cdots ,M $ based on the DoE.

 The interpolation of $ \phi_i (\boldsymbol{x})$  can be done with any surrogating technique that interpolates the data, i.e techniques where the predicted value is identical to the simulated value at the points of the DoE. 
 
Cubic spline interpolation can be used for one or two dimensional models. When considering higher dimension we can use Kriging, linear interpolation or decompose onto radial basis functions.

Starting from eigenvectors known over the DoE, a surrogate model of $\phi_i (\boldsymbol{x})$ enables us to build $\hat{\phi}_i (\boldsymbol{x}) $ $\forall \boldsymbol{x} \in D$, hence assess $\boldsymbol{\phi}_i$ over all $D$. This approach is intuitive: following the eigendecomposition, we predict the new point's coordinates with the adequate exact interpolator and as shown in Eq. (\ref{equ:equation7}) deduce the stochastic process's response.

\subsubsection{Covariance interpolation}
In this subsection a surrogate model of the covariance is used to predict the covariance  not only over the DoE points but also over the whole domain of interest $D$. For the sake of simplicity we assume that the DoE and the points where predictions are to be made add up to $M^*$ points hence $M \leq M^*$.
Let $\boldsymbol{\hat{C}}$ be the metamodel of the empirical covariance function $\boldsymbol{C}$, built using a polynomial chaos expansion for instance. $\boldsymbol{\hat{C}}$ allows one to have a predicted covariance for the $M^*$ new points of interest as follows.  
\begin{equation}
\hat{C}(\boldsymbol{x},\boldsymbol{y})= \sum_{j=0}^P a_j \psi_j(\boldsymbol{x},\boldsymbol{y}) ~~ \forall (\boldsymbol{x} , \boldsymbol{y}) \in D^2 
\label{equ:equation8}
\end{equation}

  A slightly different approach is considering an exact surrogate model, Kriging for instance. Either way, the surrogated covariance now is a $M^* \times M^*$ matrix, hence the number of eigenvectors  $\boldsymbol{\hat{\phi}}_i$ of $\boldsymbol{\hat{C}}$ is $M^*$. 

\subsubsection{ Random variables calculation } \label{Random variables}

 In the case where $H(\boldsymbol{x},\omega)$ is a Gaussian process, the  $\boldsymbol{\xi}_i$ appearing in the KL expansion in Eq.(\ref{equ:equation6}) are zero-mean, unit-variance, independant Gaussian random variables \cite{Ghanembook2003}, so no computation is needed.

When dealing with more general random processes, $\boldsymbol{\xi}_i$ are the projection of $H$ onto the base of the eigenvectors $\boldsymbol{\hat{\phi}}_i$ and given by Eq.(\ref{equ:equation5}). The integral in Eq.(\ref{equ:equation5}) cannot be calculated since $H$ is only known over the $M$ points of the DoE. To overcome this limitation, the integral is approximated with a sum involving the $M$ known values of $H$:  
 
\begin{equation}
 \hat{\xi}_i(\omega_k)=\frac{1}{\sqrt{\lambda_i}} \sum_{j=1}^{M} H(\boldsymbol{x}^{(j)},\omega_k) \hat{\phi}_i(\boldsymbol{x}^{(j)})
\label{equ:equation9}
\end{equation}

There are as many random variables $\boldsymbol{\hat{\xi}}_i$ as basis vectors $\boldsymbol{\hat{\phi}}_i$. When the eigenvectors are interpolated, the cardinality of $\boldsymbol{\hat{\phi}}_i$ and $\boldsymbol{\hat{\xi}}_i$  is $M$. In the second option (when the covariance matrix is interpolated) the cardinality of $\boldsymbol{\hat{\phi}}_i$ and $\boldsymbol{\hat{\xi}}_i$ is $M^*$.

\subsection{Flowchart of the method}
\label{Fig:method}
In the interest to simplify the approach, this section summarizes the method presented earlier in a  flowchart.
\begin{center}

\tikzstyle{block} = [rectangle, rounded corners,draw=black, thick,  fill=white,  
text centered,  text height=1 em]
\tikzstyle{line} = [draw, -latex']
\begin{tikzpicture}[node distance =  4cm, auto]
    \node [block, text width=8 cm] (a) 
    { $D= \lbrace\boldsymbol{x}^{\left( 1\right) },\dots , \boldsymbol{x}^{\left( M \right)}, \dots , \boldsymbol{x}^{\left( M^* \right)} \rbrace$ \\
    $H(\boldsymbol{x}^{\left( i\right) }, \omega_k)$ is simulated for $1 \leq i \leq M$ and for   $1 \leq k \leq N $ seeds.
    $\boldsymbol{C}$ $(M\times M)$ is the empirical covariance matrix };
    
    \node [block, below right of=a , text width=5 cm] (b) 
    {Apply an eigendecomposition to $\boldsymbol{C}$ to get the $M$ eigenvectors $\boldsymbol{\phi}_i, ~~1\leq i \leq M$};
    
    \node [block, below left of=a , text width = 4 cm] (c) 
    {Compute $\boldsymbol{\hat{C}}$, the  surogate model of the covariance operator};
    
    \node [block, below of=b , text width = 4 cm] (d) 
    {Surrogate the $M$ eigenvectors $\boldsymbol{\phi}_i$ to get $M^*$ eigenvectors $\boldsymbol{\hat{\phi}}_i,~~ 1\leq i \leq M^*$ };
    
    \node [block, below of=c ,text width = 4 cm] (e) 
    {Apply an eigendecomposition to $\boldsymbol{\hat{C}}$ to get $M^*$ eigenvectors $\boldsymbol{\hat{\phi}}_i, ~~ 1 \leq i \leq M^*$};
    
    \coordinate (Middle) at ($(e)!0.5!(d)$);
    
    \node [block, below of=Middle, text width = 5.6 cm] (f) 
    {Compute the random variables $\xi$ $$ \hat{\xi}_i(\omega_k)=\frac{1}{\sqrt{\lambda_i}} \sum_{j=1}^{M} H(\boldsymbol{x}^{(j)},\omega_k) \hat{\phi}_i(\boldsymbol{x}^{(j)})$$};

    \node (h) [block,below of=f , text width = 6.6 cm] 
    {Finaly, the KL surrogate model reads as
    $$\hat{H}(\boldsymbol{x},\omega)= \sum_i^M \xi_i (\omega)  \hat{\phi}_i (\boldsymbol{x})$$ for all $\boldsymbol{x} \in D$};   

    \path [line] (a) -- (c);
    \path [line] (a) -- (b);
    \path [line] (c) -- (e);
    \path [line] (e) -- (f);
    
    \path [line] (b) -- (d);
    \path [line] (d) -- (f);
    \path [line] (f) -- (h);
\end{tikzpicture}

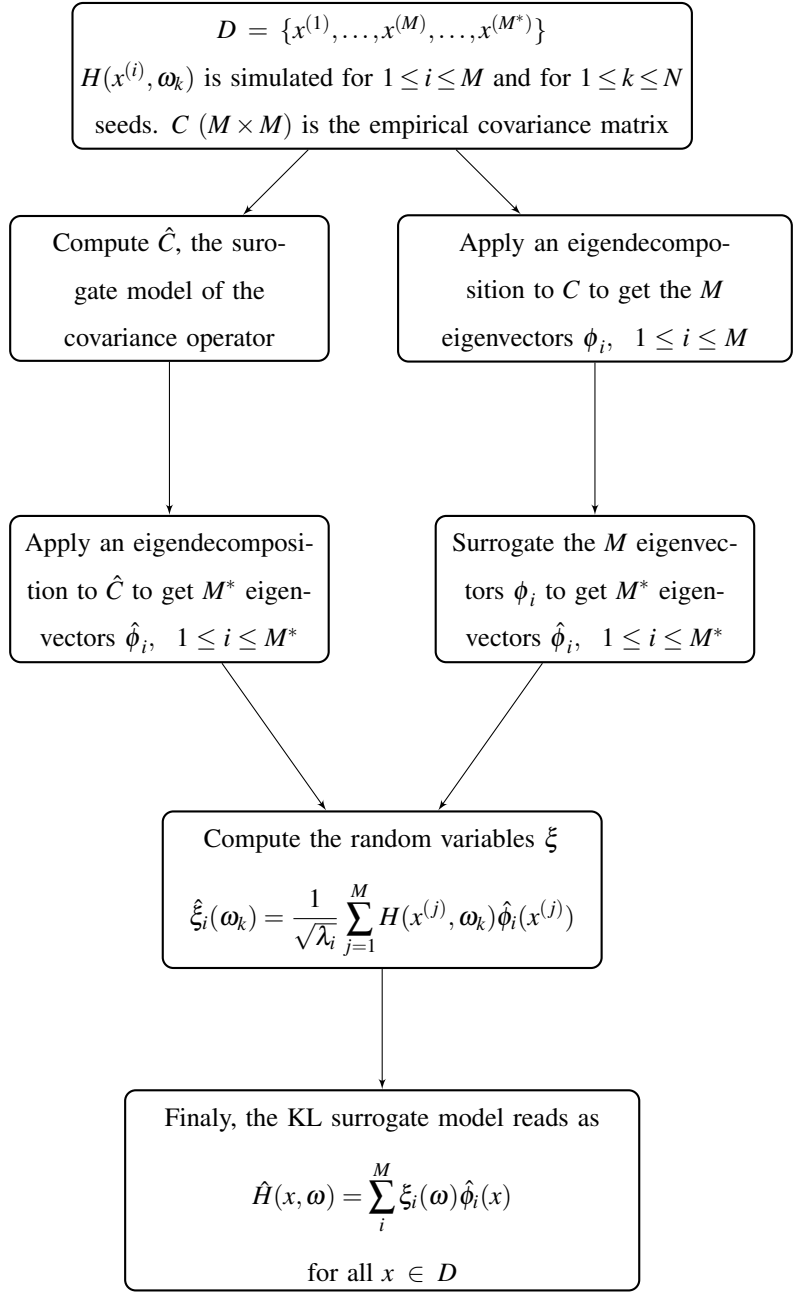
\captionof{figure}{Flowchart summarizing the method and the two possible options (surrogate modeling the covariance -right, surrogate modeling the eigenvectors -left) for building up a surrogate model of $H$.    }

\end{center}

\subsection{Error estimation and model validation}

When dealing with Gaussian processes, and since we only consider centered processes in this work, the error estimation will boil down to comparing the variance of the original stochastic simulator with that of the emulator. For a new point $\boldsymbol{x^*}$ where the surrogate is to be evaluated, one gets:

\begin{equation}
\hat{H}(\boldsymbol{x^*},\omega) = \sum_{i=1}^{p}  \sqrt{\lambda_{i}} \xi_{i}(\omega)\phi_{i}(\boldsymbol{x^*})
 \end{equation}
 where $(\xi_i, i=1, \cdots,p)$ are independent standard normal variables in this case.
Then, the associated variance reads: 
\begin{equation}
\sigma ^{2} (\boldsymbol{x^*}) = \sum_{i=1}^{p} \lambda_{i} \phi_{i}(\boldsymbol{x^*})^2  
\end{equation}
 
For non Gaussian processes, statistical tests can be applied to quantify the error of the metamodel. The Kolmogorov Smirnov (KS) test has been used to test if two drawn samples are from the same distribution (null hypothesis). The null hypothesis is rejected at level $\alpha$ if 
\begin{equation}
KS_{n,m}= \sup_{x} |F_{1,n}(x) - F_{2,m}(x)| > c(\alpha) \sqrt{\dfrac{n+m}{nm}}
\end{equation}
where $n, F_1$ and $m, F_2$ are respectively the size of the samples and their empirical distribution functions.

A more intuitive and graphical approach to compare two distributions is using histogram intersection: when it is equal to $0$, no overlap exists between the two of them, and when it is equal to $1$, they are identical. The drawback of this approach is the influence of the selection of the bins, especially for long tailed distributions. In addition to the KS test and the histogram intersection, we introduce two more metrics, namely the Hellinger distance and the Jensen-Shannon divergence.

\subsubsection{Hellinger distance}
Let $p$ and $q$ be two discrete probability measures. The Hellinger distance reads as follows: 
\begin{equation}
H(p,q)= \dfrac{1}{\sqrt{2}} ||\sqrt{p}-\sqrt{q}||_2 
\end{equation}
Hellinger distance forms a bounded ($\in [0,1]$) metric on the space of probability distribution.
\subsubsection{Jensen-Shannon divergence}
Based on the Kullback-Leibler divergence, the Jensen-Shannon (JS) divergence is a statistical method of measuring the behavior of two different distributions. A Jensen-Shannon divergence equal to $1$ indicates that the two distributions are totally different. If the Jensen-Shannon divergence is equal to $0$, the two distributions are the same almost everywhere.
We first introduce the Kullback-Leibler divergence. Let $p$ and $q$ be two discrete probability measures. Then:
\begin{equation}
 D_{KL}(p||q) = - \sum_i p(i) \log \dfrac{q(i)}{p(i)} 
 \end{equation}
Let $r=(p+q)/2$ then the Jensen-Shannon divergence reads as follow 
\begin{equation}
JSD(p,q)= \dfrac{D_{KL}(p||r) + D_{KL}(q||r)}{2}
\end{equation}

The Jensen-Shannon divergence is symmetric, finite and $0 \leq JSD(p,q) \leq 1$.

The different error metrics stated above provide a different information on how the real and the surrogated PDFs are similar. The histogram intersection metric does not provide information about the shape-similarity of two PDFs. 
To cover up this limitation, JS divergence provides an idea on how much the compared PDFs belong to a same probability family but tend to be nondiscriminant. 

\subsubsection{Model validation}

To estimate the accuracy of the surrogate's prediction, we perform a $k$-fold cross-validation: the data is partitioned onto $k$ subsets of equal size. At each step a single subsample is retained as the validation set for testing the model, and the remaining data are used to build the surrogate model. The $k$-fold validation is repeated for several partitions of the data. The error is evaluated using the error metrics defined above, namely the KS test, the histogram intersection, the Hellinger distance and the  JS divergence.\\

\section{Application to analytical examples and path loss exponent in stochastic cities}
The methods described in the previous sections have been tested on toy examples and subsequently applied to  RF exposure using stochastic cities.
 This section presents an analytical, $3$-dimensional example followed by the stochastic city case study. We remind that we are considering only centered processes in this paper. When considering the simulated data, this is achieved by removing the empirical mean prior to any treatment. The surrogate models (PCE and Kriging) are obtained with the Matlab package UQLab \cite{MarelliUQLab2014}.
\subsection{Metamodel of three dimensional process}
To test the method proposed, a toy process has been created using a known random distribution. Through simulations on different points of the design of experiment and numerous replications, we can assess the empirical covariance of the process. Let $H$ be a random process on $D=[0,2]^3 \times \Omega$:  
\begin{align}
\begin{split}
&H(\boldsymbol{x},\omega)=100 ~ \omega_1(\dfrac{1}{10} \exp (x_1  \omega_2)  + x_2 x_3 \omega_3 ) \\
 \boldsymbol{x}=(x_1,&x_2,x_3) \in [0,2]^3 
 \text{ and }
 \omega_1 \sim \mathcal{N}(0,1),~  \omega_2 \sim \mathcal{U}([1,2]),~ \omega_3 \sim \mathcal{U}([0,1])
 \end{split} 
 \end{align}

  Based on a  Latin hypercube sampling (LHS), the design of experiment (DoE) is 30 points in $[0,2]^3$, and $50$ realizations on each point, which makes a total computational cost of $1,500$ calls to the random function. The trajectories are the same for all  $30$ points of the DoE. The empirical covariance is  $C(\boldsymbol{x},\boldsymbol{y})=\mathbb{E}(H(\boldsymbol{x},\omega)H(\boldsymbol{y},\omega))$.

Following the simulations and the covariance computation, two options are tested (Figure \ref{Fig:method}). 
In the first approach we interpolate the basis vectors independently using linear interpolation at first, then using Kriging. The aim is to test the impact of  the interpolation technique on the process surrogate, hence the choice of linear metamodel ('basic' interpolator) and Kriging metamodel ('advanced' interpolator).

For the second approach a PCE surrogate model $ \hat{C}$ is built  to surrogate the covariance function:

\begin{equation}
(x_1,x_2,x_3,y_1,y_2,y_3) \in [0,2]^6 \rightarrow \hat{C}(x_1,x_2,x_3,y_1,y_2,y_3) \in \mathbb{R}
\label{equ:equation10}
\end{equation}

\begin{figure}
\centering
\begin{subfigure}[b]{0.55\textwidth}
   \includegraphics[width=1\linewidth]{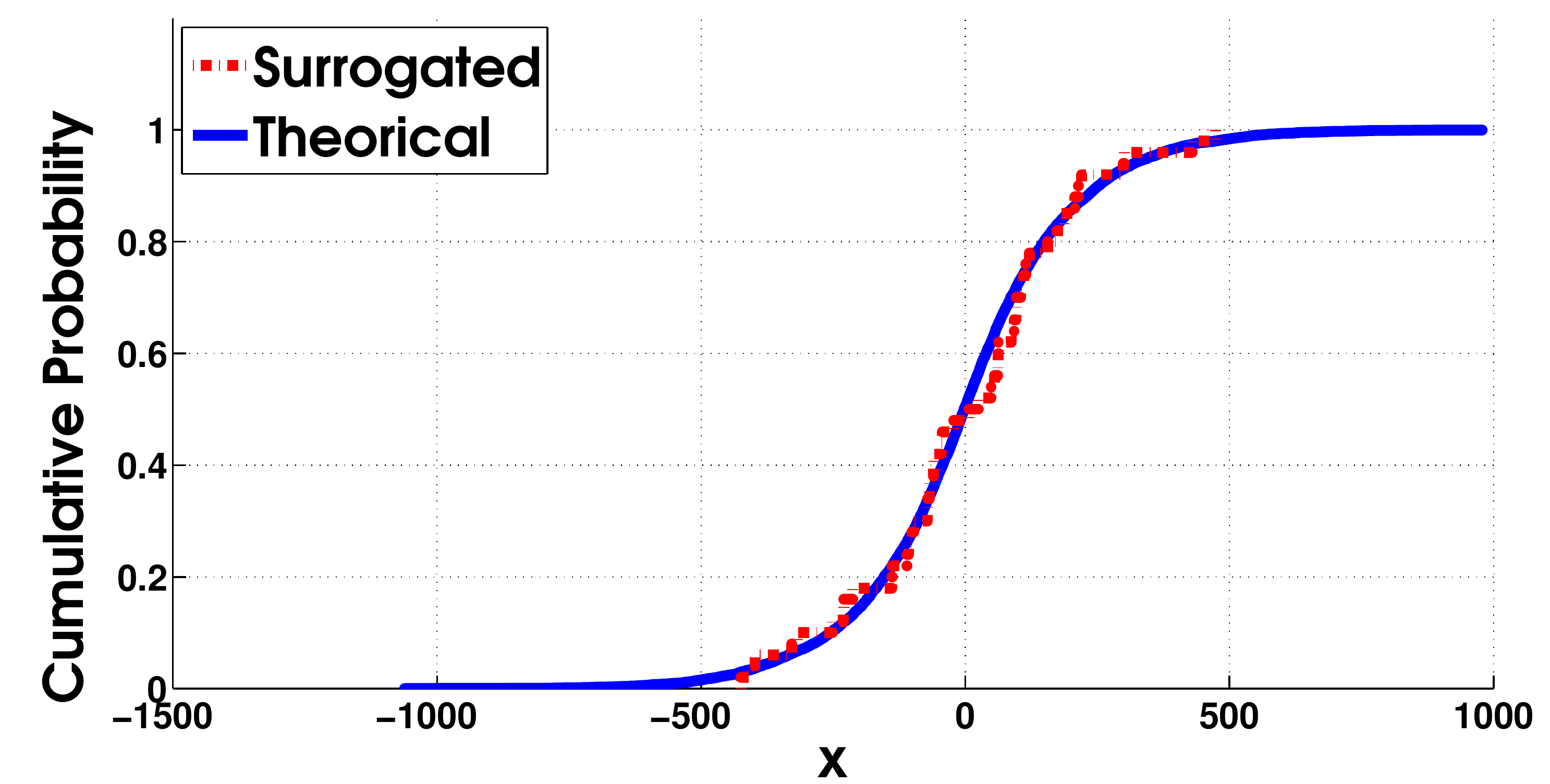}
   \caption{Linear interpolation of eigenvectors (mean histogram intersection is equal to $0.89$)}
   \label{fig:Ng1} 
\end{subfigure}

\begin{subfigure}[b]{0.55\textwidth}
   \includegraphics[width=1\linewidth]{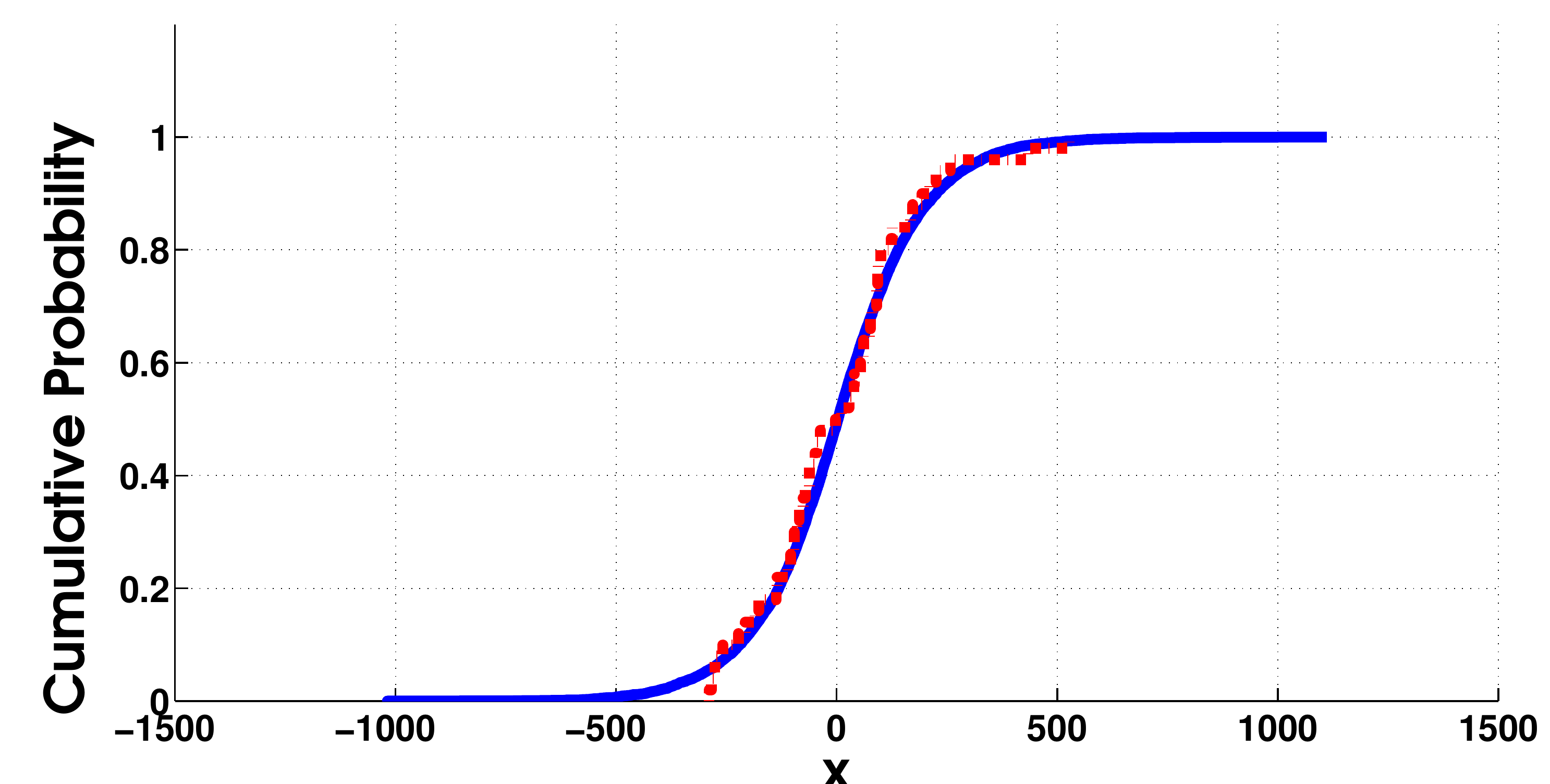}
   \caption{Kriging surrogate of the eigenvectors (mean histogram intersection is equal to $0.96$)}
   \label{fig:Ng2}
\end{subfigure}
\begin{subfigure}[b]{0.55\textwidth}
   \includegraphics[width=1\linewidth]{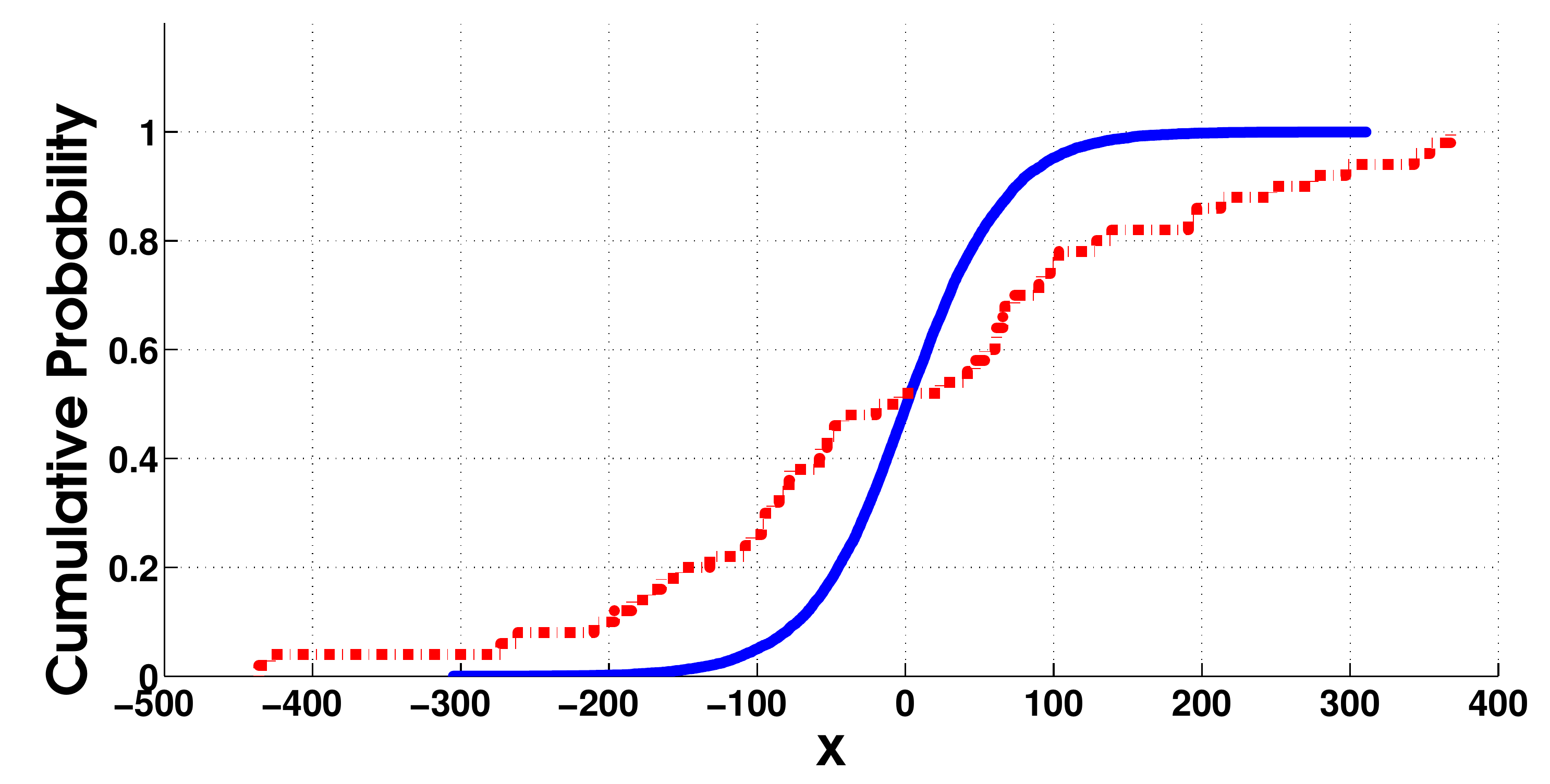}
   \caption{PCE surrogate of the covariance (mean histogram intersection is equal to $0.55$)}
   \label{fig:Ng3}
\end{subfigure}
\caption{Surrogated and true CDFs plotted in the three approaches.}
\label{fig:pdf_expl1}
\end{figure}

The covariance metamodel has $2\times 3 = 6$ inputs, and has a training set of  size  up to $29\times 29$, depending on the size of the test set. Results from both approaches are presented in Table \ref{Tab:results_expl1}. The mean value of the three error metrics evaluated over $3,000$ test points shows that surrogating the eigenvectors using Kriging performs best for this toy example. Three examples are plotted in Figure \ref{fig:pdf_expl1}, the surrogated density is computed respectively by interpolating the eigenvectors using linear model, interpolating the eigenvectors using Kriging and finally interpolating the covariance using PCE (Figure \ref{Fig:method}). The histogram intersection error in the three cases is respectively $0.89$, $0.96$ and $0.55$ (equal to the mean error (Table \ref{Tab:results_expl1})).

 \begin{table}[H]\footnotesize
\centering
\caption{Mean error over 3,000 test points.  }
  \begin{tabular}{cccc}
  \hline 
  Method & Histogram intersection & Hellinger distance & JS divergence \\ 
  \hline
  Linear interpolation of eigenvectors & 0.89 & 0.06 & 0.004 \\ 
  
  Kriging surrogate of eigenvectors & 0.96 & 0.025 & 0.001 \\ 
  
  PCE covariance surrogate & 0.55 & 0.27 & 0.03 \\ 
  \hline 
  \label{Tab:results_expl1}

  \end{tabular} 
  \end{table}

\begin{table}[H]\footnotesize
\centering
\caption{Parametric study of the error by varying the size $M$ of the DoE and the number of realizations $N$.  }
\begin{tabular}{lllllllllllll}
\hline
\multicolumn{13}{c}{$M$ the size of DoE}                                                                        \\
\cline{2-13}
& \multicolumn{3}{c}{$30$} & \multicolumn{3}{c}{$60$} & \multicolumn{3}{c}{$100$} & \multicolumn{3}{c}{$200$} \\
\cline{2-13}
& lin   & Krig  & PCE  &   lin   & Krig  & PCE & lin   & Krig  & PCE & lin   & Krig  & PCE \\
\hline
$N=50$      &$0.89$&$0.96$&$0.55$ & $0.92$ & $0.97$ &$0.56$ & $0.95$&$0.99$&$0.65$&$0.96$&$0.99$&$0.63$ \\
$N=100$     &$0.9$ &$0.96$&$0.63$ & $0.94$ & $0.98$ &$0.62$ &$0.95$ &$0.99$&$0.66$&$0.96$&$0.99$&$0.59$ \\
$N=1000$    &$0.96$&$0.98$&$0.7$  & $0.96$ & $0.99$ &$0.77$ &$0.97$ &$0.99$&$0.72$&$0.98$&$0.99$&$0.71$ \\
\hline 
 \end{tabular}
  \label{table:convergence}
\end{table}

 To characterize the dependence of the method on the surrogate model used, the size of input data $M$ and the  number of realizations, the histogram intersection error is estimated, and results are presented in Table \ref{table:convergence}. For this comparison, only the histogram intersection metric is used. Hellinger distance varies in the same way as the histogram intersection and JS divergence did not seem to be discriminant.
 
 For the examples tested, the ranking of the three approaches depends on the process, i.e. for the first example the Kriging is always performing better $\forall M \in \mathbb{N},~ \forall N  \in \mathbb{N}$. Meanwhile for the second example (Section \ref{Seq:exple2}), the ranking is the other way around, meaning that the PCE performs better than the other options for $M=30,~ N=50$. Table \ref{table:convergence} shows that the performance increases when $M$ and/or $N$ increases. That said, increasing $N$ seems to grant a better accuracy compared with increasing $M$.

  The poor performance of the PCE surrogate points out to the dependence of the overall method on the eigenvectors and their computation and is probably due to the following reasons:
  \begin{itemize}
  \item For a data set of size $M=30$, there is $30 \times 30 = 900 $ covariance terms. Hence the surrogate model of the covariance will have $900$ inputs (as in Eq. \ref{equ:equation10}), the PCE model might get noisy and over-fitted.  
  \item The covariance surrogate $\boldsymbol{\hat{C}}$ has been symmetrized, meaning that if the obtained metamodel of the covariance is denoted $C_*$ (which is not necessarily a symmetric function of its inputs $(\boldsymbol{x},\boldsymbol{y})$) then we use $ \boldsymbol{\hat{C}} = \dfrac{\boldsymbol{C}_* + \boldsymbol{C}^{\intercal}_*}{2}$. This step may contribute to the noisy results.  Surrogate modeling $\boldsymbol{C}$ only on a triangular domain has been tested, yet the performance on the same test points did not improve.  
  \end{itemize}

 The error is always evaluated  between the simulated and the surrogated PDF (using one of the three options), mainly because in case studies the real PDF in usually unknown, hence comparing the surrogate and the original simulator is impossible. \\

\subsection{Metamodel of the path loss exponent distribution}
\label{Seq:exple2}
Using the stochastic city simulator, $8\times 10^5$ rays have been generated and launched from an $30m$ high antenna in a city measuring $425~m^2$ and fully determined by a seed number and three input variables detailed in Table  \ref{table:table1}.
 
The seed (from $1$ to $50$) is used to initialize a pseudo-random number generator in the stochastic city generator. These seeds are used to freeze the trajectory of the process in the sense that two cities with the same seed number and the same parameters are exactly the same, and accordingly their path loss exponents are identical.\\
\begin{table}[H]

\centering
\caption{Input variables for the stochastic city generator. }
 \begin{tabular}{ l c }
 \hline
 Input parameter & Range \\
 \hline 
 Street width $x_1$ & $[10~m,20~m]$ \\ 
 
 Building height $x_2$  & $[9~m,18~m]$ \\ 
  
 Anisotropy $x_3$ & $[0,1]$ \\ 
 \hline 
 \end{tabular} 
 
 \label{table:table1}
 \end{table} 
The DoE is a LHS of $30$ cities for the $50$ seeds (meaning in total $30 \times 50=1,500$ simulations) with $10 \%$ of the data for testing.

 For both examples a $k$-fold cross validation was carried out by dividing the data ($30$ points) onto $k=10$ subsets. At each step a surrogate model of the stochastic simulator is built using nine out of the ten subsets. The remaining subset is used to evaluate the performance of the model.  This procedure is then repeated for  $100$ different partitions of the data set.

\begin{figure}
\centering
\begin{subfigure}[b]{0.55\textwidth}
   \includegraphics[width=1\linewidth]{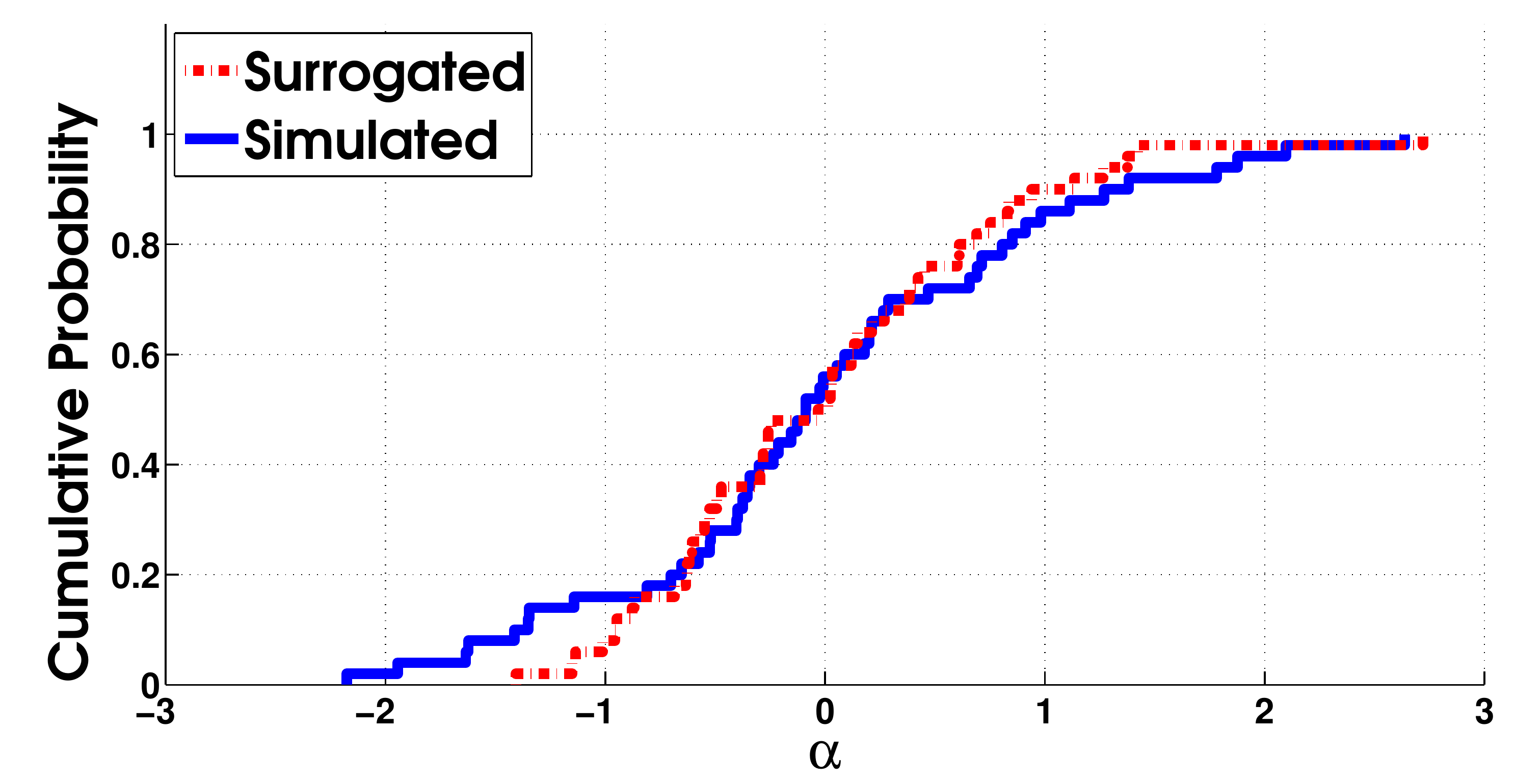}
   \caption{Linear interpolation of eigenvectors (mean histogram intersection is equal to $0.74$)}
   \label{fig:Ng1} 
\end{subfigure}

\begin{subfigure}[b]{0.55\textwidth}
   \includegraphics[width=1\linewidth]{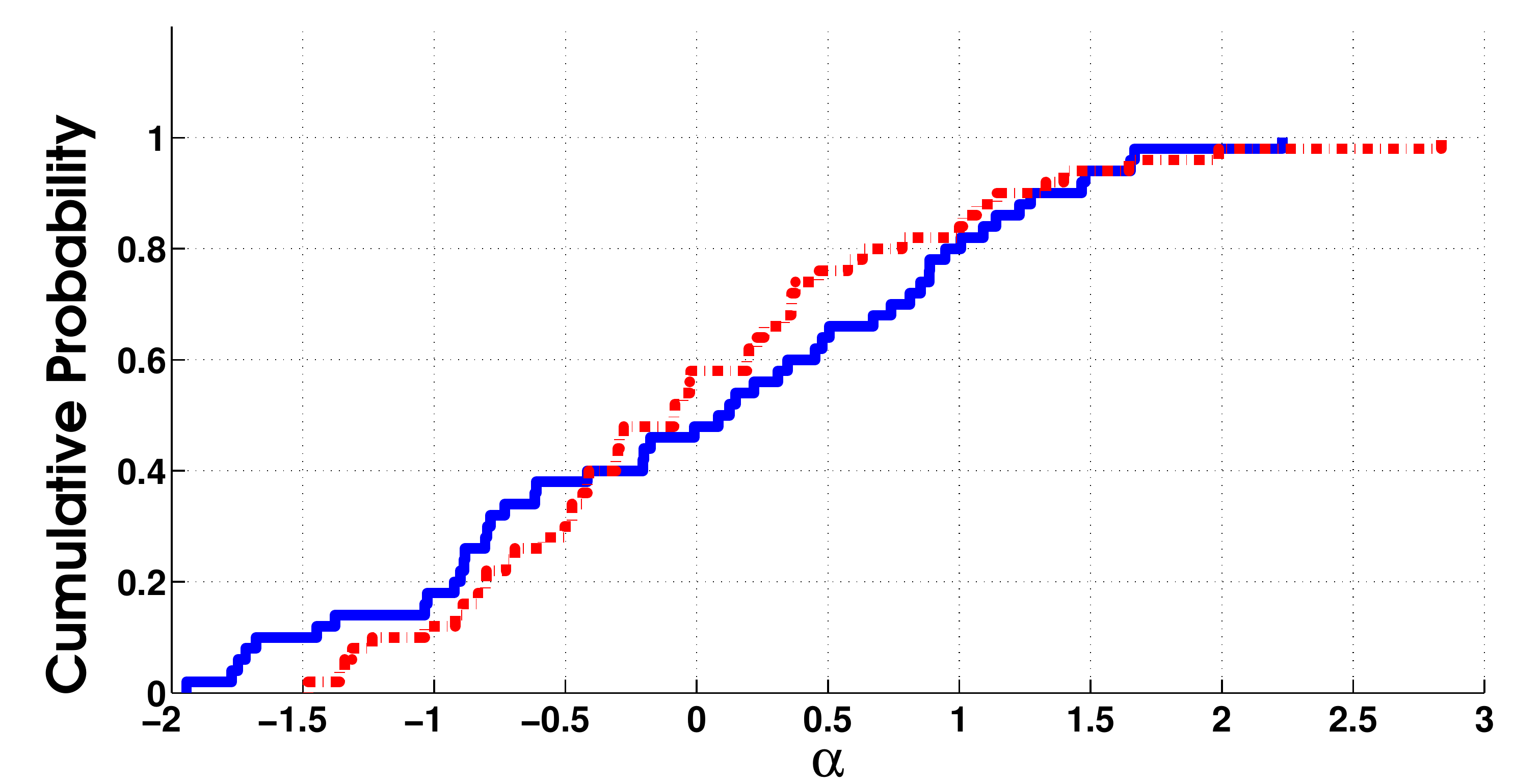}
   \caption{Kriging surrogate of the eigenvectors (mean histogram intersection is equal to $0.71$)}
   \label{fig:Ng2}
\end{subfigure}
\begin{subfigure}[b]{0.55\textwidth}
   \includegraphics[width=1\linewidth]{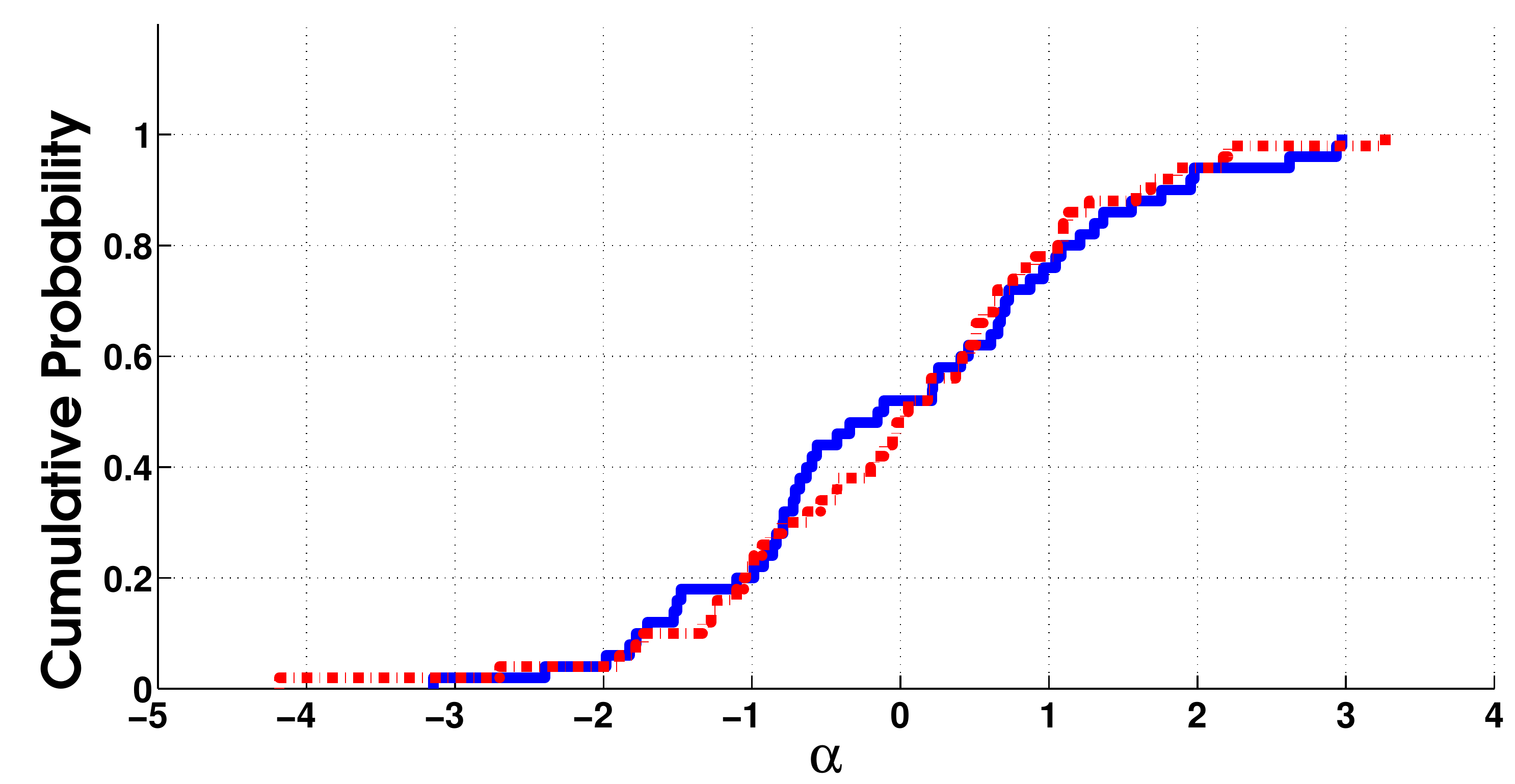}
   \caption{PCE surrogate of the covariance (mean histogram intersection is equal to $0.76$)}
   \label{fig:Ng3}
\end{subfigure}
\caption{Surrogated and simulated CDFs plotted in the three approaches.}
\end{figure}

 \begin{table}[H]\footnotesize
\centering
\caption{Mean error over 3,000 test points.}
\centering
  \begin{tabular}{cccc}
  \hline 
  Method & Histogram intersection & Hellinger distance & JS divergence \\ 
  \hline
  Linear interpolator of eigenvectors & 0.74 & 0.15 & 0.02 \\ 
  
  Kriging surrogate of eigenvectors & 0.71 & 0.17 & 0.03 \\ 
  
  PCE covariance surrogate & 0.76 & 0.14 & 0.02 \\ 
  \hline 
  \end{tabular}
   
  \end{table}

For this example, the dependence of the method on $M$ and $N$ could not be evaluated since only $30$ points were simulated (due to the high computational cost).

In this example KS test is used to test the null hypothesis: the predicted and the sampled PDFs come from the same distribution. Results are $4.8\%$, $12.6 \%$, $1.46\% $ of rejection of the null hypothesis (at level $ 5\%$) for the respectively linear surrogate of $\boldsymbol{\phi}_i$, the Kriging surrogate of $\boldsymbol{\phi}_i$, and the PCE surrogate of the covariance $\boldsymbol{C}$.

The KS test allows one to rank these three approaches that none of the three error metrics could provide since they all showed that the performance of the three approaches is more or less the same. Considering the fact that the size of the training set was small because of the high computational cost of the original stochastic simulator, the error has been considered acceptable.\\

\section{Conclusion}

This study describes a non parametric surrogate model of stochastic simulators based on Karhunen-Lo\`eve (KL) expansion. The approach has been tested first on closed-form processes in order to validate the method, and after that applied to a full scale problem linked to the assessment of a population exposure induced by base station antennas.

 The eigenvectors of the KL expansion in the domain of interest  has been predicted in two different ways : at first a surrogate modeling of the process covariance operator using polynomial chaos expansions (PCE) has been used. The second approach, consists in directly surrogating the eigenvector. In terms of performance, the error evaluation on the toy example  shows better results when the eigenvectors are surrogated using the Kriging. Nevertheless, for the path loss exponent (PLE) example, the PCE metamodel of the covariance performed better.

 For both examples, and when the eigenvectors are interpolated using either Kriging or a linear interpolator, the tests performed do not show a significant difference in the overall performance. This is mainly due to the multiple steps governing the stochastic metamodeling procedure. Hence the eigenvector interpolation error fades away into the global error. Nonetheless the first example shows that the empirical covariance and its eigenvectors play a crucial role in the precision of the decomposition (the PCE surrogate performed poorly).

 Considering the error, the size of the DoE $M$, and the number of realizations $N$, impact the accuracy of the covariance matrix and the precision of its surrogate but also the accuracy of the random variables appearing in the KL expansion. 
  The central limit theorem can be used to evaluate the error of the covariance matrix, but once the covariance or its eigenvectors are surrogated, we lose track of the analytical error, since errors from the surrogate model of the covariance, its parameters and the sampling over $M$ points were added.

The fact that the randomness in the case study was 'controllable' (through freezing the same seed $\omega_k$ for different points of the DoE) is a key characteristic, since it enabled us to compute all the terms of the expansion. 

Surrogate modeling stochastic simulators using non parametric approaches requires a large number of simulations, for instance, 1,500 simulations were needed for the PLE example (3 inputs). The parsimony aspect of the approach requires further investigation.

 \section*{Acknowledgment}
This paper reports work undertaken partially in the context of the ANSES project AMPERE (ANSES PNR EST-2016-2 RF-04).


\bibliography{AZZI_Stocha} 

\end{document}